\documentclass[graybox]{svmult}

\usepackage{mathptmx}       
\usepackage{helvet}         
\usepackage{courier}        
\usepackage{makeidx}         
\usepackage{graphicx}        
\usepackage{multicol}        
\usepackage[bottom]{footmisc}

\makeindex             

\begin{document}

\title*{Symmetry and Universality in Language Change}
\author{Richard A Blythe}
\institute{Richard A Blythe \at SUPA, School of Physics and Astronomy, University of Edinburgh, Peter Guthrie Tait Road, Edinburgh EH9 3FD, UK \email{R.A.Blythe@ed.ac.uk}}
%
%
\maketitle

\abstract*{We investigate mechanisms for language change within a framework where an unconventional signal for a meaning is first innovated, and then subsequently propagated through a speech community. We appeal to the notion of universality as it applies to complex interacting systems in the physical sciences and which establishes a link between generic (`universal') patterns at the macroscopic scale and relates them to symmetries at the microscopic scale. By relating the presence and absence of specific symmetries to fundamentally distinct mechanisms for language change at the level of individual speakers and speech acts, we are able to draw conclusions about which of these underlying mechanisms are most likely to be responsible for the changes that actually occur. Since these mechanisms are typically believed to be common to all speakers in all speech communities, this provides a means to relate universals in individual behaviour to language universals.}

\abstract{We investigate mechanisms for language change within a framework where an unconventional signal for a meaning is first innovated, and then subsequently propagated through a speech community to replace the existing convention. We appeal to the notion of universality as it applies to complex interacting systems in the physical sciences and which establishes a link between generic (`universal') patterns at the macroscopic scale and relates them to symmetries at the microscopic scale. By relating the presence and absence of specific symmetries to fundamentally distinct mechanisms for language change at the level of individual speakers and speech acts, we are able to draw conclusions about which of these underlying mechanisms are most likely to be responsible for the changes that actually occur. Since these mechanisms are typically believed to be common to all speakers in all speech communities, this provides a means to relate universals in individual behaviour to language universals.}

\section{Three Notions of Universality in Language Change}
\label{sec:intro}

Language is a system of behaviour that is acquired by social learning, that is, by learning from other members of a social group as opposed to a process of individual exploration \cite{boy85}. On the face of it, the social interactions where a linguistic behaviour is transmitted from on individual to another are highly specific. Each interaction could depend on the the goals of the participants in the interaction, their own individual history of usage, the relative social standing of the individuals involved, to name just three factors that have been discussed in the literature \cite{kel94,cro00,lab01}.  Nevertheless, when one looks at the system that arises from these repeated social interactions, common patterns emerge.

Some of these patterns relate to the structure of language itself. For example, typological surveys show that although six different orderings of the subject (S), verb (V) and object (O) are possible, two particular orderings (SOV and SVO) are much more common than any of the others (see Feature 81A in \cite{wals13}). Other patterns relate to how languages change over time, in particular those cases where one conventional signal for a meaning is replaced by another \cite{cro00}. A number of these linguistic patterns are surveyed in \cite{tag11}. These include the `male lag', which relates to the common observation that when a change is in progress, it is the females who lead the change (i.e., are less likely to be users of the outgoing convention). Meanwhile, when partitioning language users by age, rather than gender, one typically encounters an `adolescent peak', whereby the age group leading the change is not the very youngest, but the adolescents. Finally, the frequency of the new convention as a function of time tends to follow an S-curve, that is starting slowly, then accelerating, before tailing off as the old convention is eliminated. Indeed, this pattern is seen not only in language change, but also in other types of cultural change, such as the adoption of a technological innovation \cite{rog03}.  All of the phenomena described in this paragraph might be described as \emph{universal}, in the sense that they have been observed in different social groups at different times, and in some cases even across more than one type of cultural behaviour.

This however is not the only possible notion of universality that relates to language change (or cultural evolution more generally). The linguistic behaviour that is displayed and transmitted in social interactions is determined to some extent by the cognitive and physical apparatus possessed by the interacting agents. For example, in the case of word order, it is possible that sentences that have the subject first are easier for humans to process than other types of sentence, which would be expected to lead to those subject-first sentences being more common across the world's languages. A variety of such linguistic principles have been proposed: see e.g.~\cite{mau11} for a discussion in a psychological context. Likewise, articulatory or auditory constraints may cause certain vocalisations to be more easily produced or understood than others \cite{oha83}. The crucial point is that these constraints are assumed to be common to all language users, no matter which social group they belong to: in this sense (and one that is distinct to the above) these abilities are \emph{universal}.

It is natural to expect some sort of link between these two types of universals: that is, to propose that the origin of universal patterns of cultural evolution lies in the universal constraints that underpin the social interactions and social learning.  What is unclear is whether the relationship is simple and transparent. In this case, every phenomenon that is seen at the macroscale would be directly observable in individual interactions. On the other hand, the relationship might be rather more complex, arising from multiple biases and the fact that the behaviour has been acquired and reproduced multiple times.  Experimental work provides evidence in favour of both positions (e.g., \cite{cul12,rea09}), which is perhaps not surprising since they are not mutually exclusive.

One tool that is becoming increasingly widely used to understand the link between universals at the individual and population level is mathematical modelling of complex interacting agent systems \cite{cas09,hru09,smi14}. Here, a great deal of intuition is drawn from the experience of modelling physical systems of interacting particles (atoms and molecules) that collectively form macroscopic structured materials (for example, metals). In this context, one encounters a notion of \emph{universality} that is again distinct to the two cited above.  Roughly speaking, this notion pertains to the link between individual-level and collective behaviour, and in this work we shall draw inspiration from it to understand how the way in which a language change propagates can be related to individual behaviour.

It is instructive to discuss briefly a concrete example of universality in condensed matter physics to elucidate our approach.  Magnetic materials are characterised by a \emph{Curie temperature}, below which they exhibit permanent magnetism \cite{kit05}.  For iron, the Curie temperature is $770^\circ{\rm C}$, which is why an iron bar serves as a good choice for a bar magnet in a child's chemistry set\footnote{One may ask what a magnet is doing in a `chemistry set', given that magnetism is physics, but this is beyond the scope of this article.}.  At a distance $\Delta T$ below the Curie temperature, the strength of the magnet increases as a power law $(\Delta T)^\beta$ (at least in the range where $\Delta T$ is small) \cite{gol92}.  It is this exponent $\beta$ that is universal: it has the same numerical value for a wide variety of magnets with different microscopic structures.  For example, model magnets whose component parts interact with different strengths or different ranges, or have different spatial arrangements, all have the same power-law exponent $\beta$ \cite{gol92}.

We can now state more precisely what is meant by universality in this context. It applies when some macroscopic phenomenon is observed independently of the details of the interactions between the component parts \emph{as long as} these interactions are consistent with a certain set of general principles.  In condensed matter physics, these principles relate to the \emph{symmetry} of the system \cite{gol92}.  In the example of the magnet, the relevant symmetry property is that the interactions are unaffected if one exchanges all north and south poles of the microscopic magnets that collectively form the macroscopic magnet.

In the remainder of this article, we will see how similar ideas relating to symmetry in linguistic interactions between speakers can be used both to predict the emergent dynamics of language change, and to categorise different theories for the factors that may influence individual behaviour.  As we will see, various types of asymmetry are possible, and each corresponds to a characteristic pattern of language change, only some of which are consistent with the universal S-curve of language change mentioned above (and discussed in further detail below).  While the main results outlined here were established in the context of a specific model in Ref.~\cite{bly12}, we offer here a much broader perspective than was achieved in this earlier work. In particular, we present some new general results that apply to a wide range of models that respect the relevant symmetries while differing in detail.  As such, they underline the utility of considerations based on symmetry as a means to understand the behaviour of complex interacting systems outside the physical sciences.

\section{Asymmetry in Language Change: The Universal S-Curve}

\begin{figure}[b]
\includegraphics[width=\linewidth]{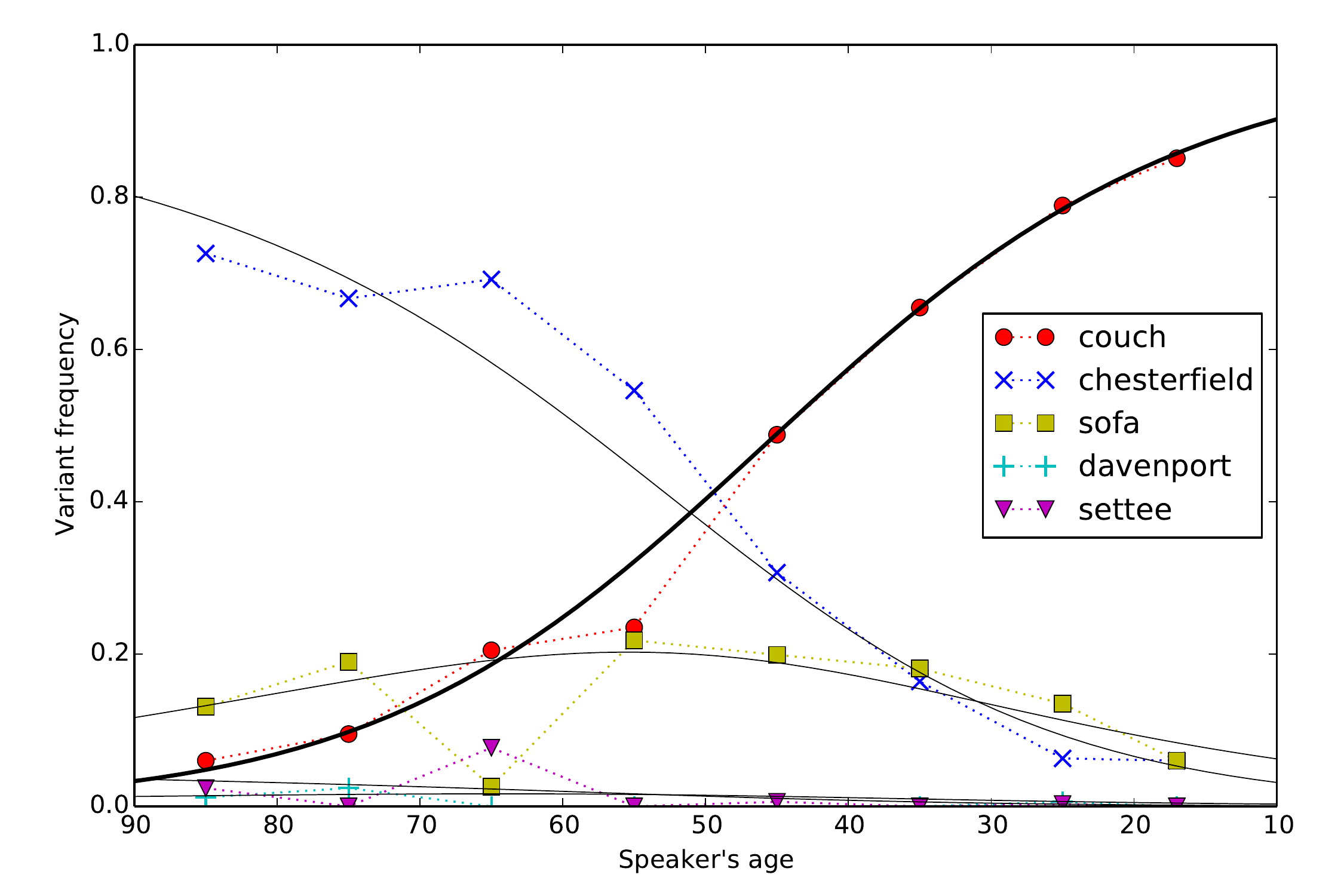}
%
%
\caption{Variation in frequency of different furniture terms in Canadian English. Points connected with dotted lines correspond to empirical data from \cite{cha95}. This was an apparent-time study, meaning that speakers of different ages were surveyed. In this framework, older speakers are assumed to be representative of typical behaviour at an appropriate point in the past. Hence plotting the data as a function of decreasing age gives an estimate of the real-time change trajectory for these data. Solid lines are fits to the functions (\ref{mvlog}); the thickest line is that with the largest growth rate, and is the ultimately winning variant that follows the S-curve.}
\label{fig:scurve}       
\end{figure}

We begin by stating more precisely the properties of the universal S-curve of language change, and highlight why this points towards some underlying asymmetry in the system.  Throughout this work, we will always have in mind the the case of a language change in which the conventional signal for a specific meaning is replaced over time with a new signal.  Specific examples include the marking of the future tense in Brazilian Portuguese \cite{pop07}, negation in French \cite{gri09} and the word used by English speakers in Canada to refer to the item of furniture that I (and the people I typically interact with\footnote{An exception is my three-year old child, who has elected for `couch'.}) call a `sofa' \cite{cha95}.  In each of these cases, the frequency that the incoming variant (`couch') is used follows an S-curve trajectory:  the rate of growth initially accelerates until both incoming and outgoing variant are widely used, after which the rate of growth decelerates as the incoming variant becomes established as a convention (i.e., a variant that is used by a large majority of speakers).  The empirical data for Canadian furniture terms is shown in Fig~\ref{fig:scurve}, where the rise of `couch' has been fit to an idealised logistic S-curve that has been discussed widely in the historical and sociolinguistics literature (see e.g.~\cite{gre54,kro89,cro00,cha02,den03}).

The case of Canadian furniture terms is an interesting one as multiple variants are in simultaneous competition. Among the older speakers, `sofa' and `couch' are the most used low-frequency variants, and indeed given this information, one might expect `sofa' to be the term most likely to displace the existing convention (`chesterfield'). This shows that there are (at least) two factors that determine the dynamics of one of the variants over time: its initial frequency $f_i$, and its rate of growth, $s_i$. In the absence of competition, the frequency grows as $x_i(t) = f_i {\rm e}^{s_i t}$. The effect of competition is included by ensuring that all the frequencies sum to 1: $\sum_i x_i(t) = 1$. Then,
\begin{equation}
\label{mvlog}
x_i(t) = \frac{ f_i {\rm e}^{s_i t} }{ \sum_j f_j {\rm e}^{s_j t} } \;.
\end{equation}
Depending on the initial frequencies and growth rates, one can arrive at a variety of different shapes of curve, as shown in Figure~\ref{fig:scurve}. The key point is that the variant with the largest growth rate ($s_i$) will eventually saturate to $x_i(t)=1$ and, if it starts at low frequency, will typically follow the characteristic S-shaped curve.

Another way to understand the S-curve---and in particular its symmetry prop\-erties---is to take a dynamical systems theory view.  The rate of change of the variant frequency when it is close to $0\%$ and $100\%$ is sufficiently small that it can be idealised to zero.  This implies that these are fixed points of the dynamics.  However, the initial state is an unstable fixed point (repulsive) while the final state is stable (attractive).  Thus there is an asymmetry in the stability of these two fixed points, which in turns points towards some underlying asymmetry in the system of linguistically-interacting agents.  As we will see in the following, there are a number of ways in which this asymmetry may be generated: however, not all of them are equivalent in terms of the language change trajectories that arise.

We emphasise that our paradigm throughout this work is the case where an existing convention is being replaced by an innovative variant signal for the same meaning.  The innovation process itself is not directly modelled; rather, it is implicit in the initial condition, which will be a very low (but nonzero) frequency for the innovation. 

\section{Language Change with No Asymmetry}
\label{sec:neutral}

For orientation, we ask the following question: What would language change look like if there are no asymmetries at all?  This is a very strong requirement.  First, every member of a speech community must behave identically. Every group of speakers that interacts---be this in pairs, triads or larger units---must interact with the same frequency, and each speaker must react in the same way to the behaviour of the speakers they interact with.  They must also give no preference to any of the variants (e.g., different words for `sofa') over any other that they are exposed to. This already shows that there are at least three ways to generate asymmetry, and we shall consider them all below.

Suppose the innovation (incoming variant) is used in the speech community with a frequency $x$. Here, by \emph{frequency} we mean a number between $0$ and $1$ which corresponds to the fraction of utterances where a specific meaning is being expressed in which the innovative signal is used. By implication, we have that the convention has a frequency $1-x$.

If no asymmetries are allowed, then all speakers can do is produce each variant in proportion to what they have heard.  This means that the expected rate of change of any variant is zero: that is, the innovation frequency $x$ is just as likely to go up as to go down in any time interval.

We can demonstrate this result by appealing a fairly general mathematical model (and one that we will modify in later sections to explore the link between symmetry and the resulting language change process). Let $G_{ij}$ be the probability that agents $i$ and $j$ interact in a time interval lasting $\delta t$. The frequency of the innovation experienced by speaker $i$ over this time interval is 
\begin{equation}
\label{eq:xitilde}
\tilde{x}_i = \frac{\sum_{j\ne i} G_{ij} x_j}{\sum_{j\ne i} G_{ij}} \;.
\end{equation}
The quantity appearing in the denominator here, $G_i = \sum_{j\ne i} G_{ij}$ is the total probability that agent $i$ interacts with another speaker in the time interval $\delta t$. 

Now, if speaker $i$ participates in an interaction in this time interval (an event that occurs with probability $G_i$), then we suppose that it updates its usage frequency to equal some average of its existing value $x_i$ and the frequency $\tilde{x}_i$ observed in the interaction.  Otherwise, if it does not interact (probability $1-G_i$), the usage frequency remains unchanged. That is, the mean value of a speaker $i$'s usage frequency after an interaction, $x_i'$, is
\begin{equation}
\label{eq:combine}
x_i' =  G_i[ \alpha x_i + (1-\alpha) \tilde{x}_i] + (1-G_i) x_i = x_i + G_i(1-\alpha) ( \tilde{x}_i - x_i)
\end{equation}
where $\alpha$ is a number between $0$ and $1$ that specifies how resistant a speaker is to change. In the case $\alpha=0$, a speaker immediately accommodates to the usage frequency of its interlocutors; in the case $\alpha=1$ it never changes. Since there are no asymmetries, $\alpha$ is the same for all speakers, and we avoid the pathological (and uninteresting) case of no change,
 $\alpha=1$.

We can now work out what the overall frequency of the innovation is after all speakers have updated their individual frequencies.  We find
\begin{equation}
\label{eq:xdash1}
x' = \frac{1}{N} \sum_i x_i' = \frac{1}{N} \sum_i \left[ x_i + G_i(1-\alpha)(\tilde{x}_i - x_i) \right] = x + \frac{1-\alpha}{N} \sum_{i} G_i(\tilde{x}_i - x_i)
\end{equation}
where $N$ is the number of speakers.  Notices that strictly speaking what we have calculated here is the \emph{mean} frequency of the innovation in the population, where the average is over all possible interactions that might happen in the time interval $\delta t$.  For large speech communities, we are justified in ignoring fluctuations which are expected to be of order $1/\sqrt{N}$ (unless we find that the expected changes in $x$ are themselves of similarly small magnitude).

Now, symmetry demands that $G_{ij} = G_{ji}$---the frequency that $i$ interacts with $j$ equals the frequency that $j$ interacts with $i$. (Actually, this will always be true, although the response to the interaction need not be symmetric, as discussed below).  This has the following important consequence:
\begin{equation}
\label{eq:cancel}
\sum_i G_i \tilde{x}_i = \sum_i \sum_{j\ne i} G_{ij} x_j = \sum_j x_j \sum_{i\ne j} G_{ij} =  \sum_j x_j \sum_{i\ne j} G_{ji} = \sum_j G_j x_j = \sum_i G_i x_i \;.
\end{equation}
Using this in (\ref{eq:xdash1}), we find
\begin{equation}
\label{eq:xdash2}
x' = x + \frac{1-\alpha}{N} \left[ \sum_i G_i \tilde{x}_i - \sum_i G_i x_i \right] = 
x + \frac{1-\alpha}{N} \left[ \sum_i G_i x_i - \sum_i G_i x_i \right] = x \;.
\end{equation}
This shows that the expected frequency of the innovation in the speech community, $x'$, at the end of a time interval lasting $\delta t$ is the same as its value, $x$, at the start of that time interval. In other words, variant frequencies do not change on average.

\begin{figure}[b]
\includegraphics[width=\linewidth]{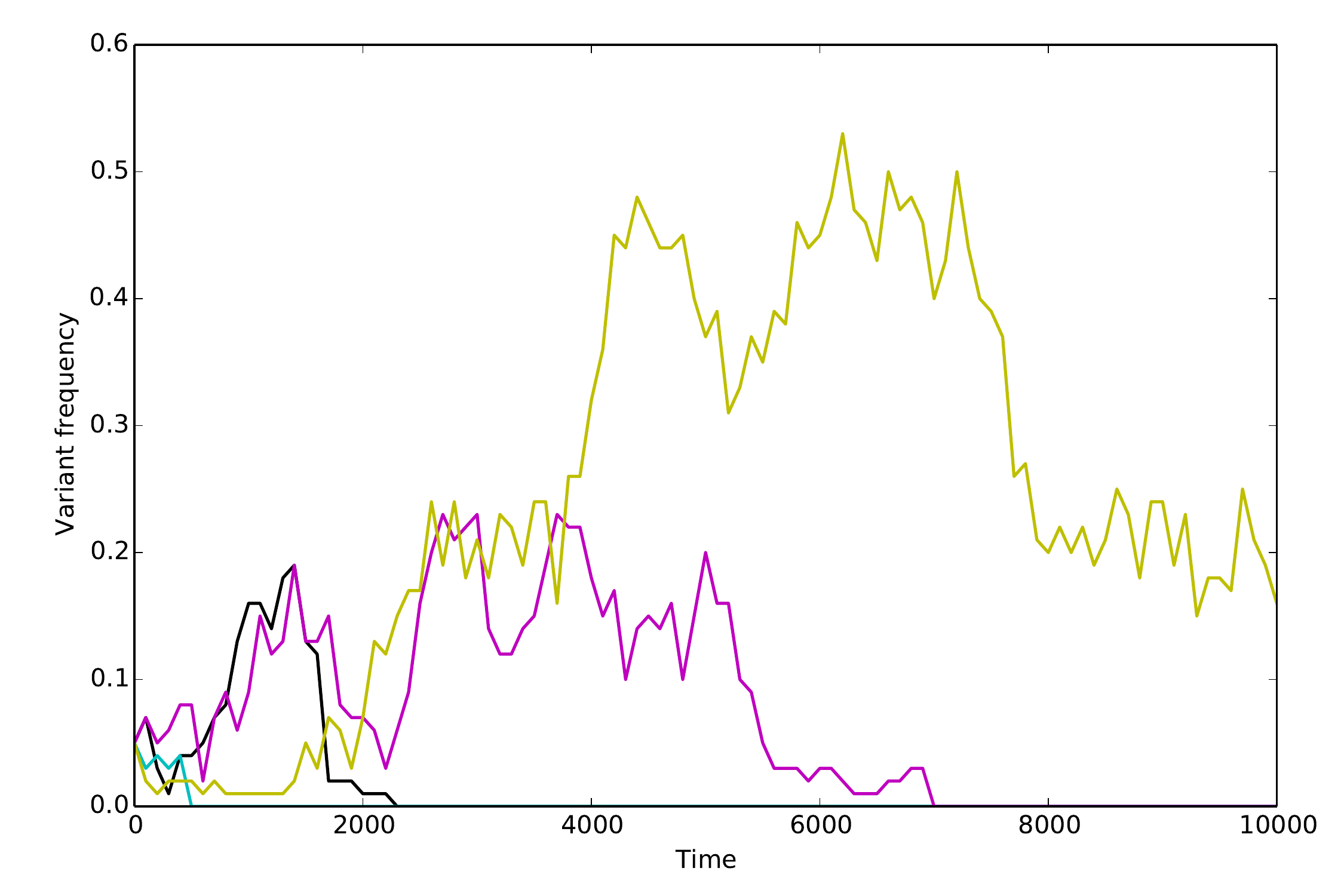}
%
%
\caption{Four change trajectories generated by genetic drift with an initial frequency of the innovation of $0.05$. In each case, strong upward and downward fluctuations are observed, and no directed S-curve change trajectory (similar to that seen in Fig.~\ref{fig:scurve}) is seen. Three of the innovations go extinct (one rather quickly); one is still present at the end of the time period shown.}
\label{fig:drift}       
\end{figure}

At this point it is necessary to return to the observation that there are fluctuations of order $1/\sqrt{N}$ around this average change in frequency.  That is, in any real system we will expect to see small changes in variant frequencies from one time step to the next due to fluctuations in the identities of the speakers who interact, and their response to the interaction. However, the symmetries inherent in these interactions imply that the probabilities probabilities of upward and downward fluctuations are the same.  Consequently, the small fluctuations in variant frequencies are undirected.  Given enough time, it is possible for one of the variants to be eliminated by chance, at which point it will not (at least in the class of models under consideration here) be reinvented.  The canonical mathematical model for this random processes with these characteristics is genetic drift that was introduced mathematically in the 1930s \cite{fis30,wri31}.  Typical trajectories of change generated by genetic drift are shown in Fig.~\ref{fig:drift}, and can be seen to differ significantly from the directed S-curve of Fig.~\ref{fig:scurve}. In particular, both fixed points (at $x=0$ and $x=1)$ are stable, as one would expect if there is no underlying asymmetry.

For what follows, it is perhaps worth emphasising the symmetries assumed in this analysis. First, all possible variants are considered equivalent. Further, all speakers are equivalent in terms of their propensity to change (all have the same $\alpha$ value). Dyads are symmetric: when speaker $i$ interacts with speaker $j$, speaker $j$ interacts with speaker $i$ (and they interact in the same way). More subtly, we assumed that a speaker's updated usage frequency would be some linear combination of its existing frequency and those if its interlocutors.  As we will see below, nonlinear functions correspond to distinguishing between variants by their usage frequencies. A fully symmetric model would preclude giving a higher (or lower) weight to a more frequent variant. We explore the effect of relaxing each of these symmetries in the following sections.

\section{Asymmetry in Interaction Frequencies}

A fully symmetric model would have all interaction frequencies $G_{ij}$ equal.  Note, however, that we did not make this assumption in the previous section.  Therefore, the main conclusion---that a variant's frequency exhibits undirected fluctuations---should hold for arbitrary variation in $G_{ij}$ between pairs of speakers.  It turns out that this is indeed the case, as was shown quite generally for a wide range of evolutionary processes (including cultural evolution) on complex network structures \cite{bax08}.  In fact, this insensitivity to network structure in the dynamics goes much deeper, in that the fluctuations in the usage frequency $x_v$ at the community level depend only on the size of the speech community, and not on the details of who speaks to whom and how often \cite{bax08,bly10}. This finding is reminiscent of the concept of universality as it applies to magnets, where the spatial arrangement of atoms in the solid did not affect its overall magnetic properties (see Section~\ref{sec:intro} above).

It would be somewhat unreasonable to expect every member of a speech community to have the same number of interlocutors, and to interact with each of them with exactly equal frequency. Consequently, the fully symmetric theory of the previous section has never (to my knowledge) been advanced as a linguistic theory for language change.  However, the extension to the case where interaction frequencies can vary and speakers adopt a some average of their own and their interlocutor's frequencies for future interactions \emph{has} been advanced as the linguistic theory of \emph{determinism}, at least as it applied to new-dialect formation under certain circumstances \cite{tru04}.  The psychological basis for this theory is accommodation, a process whereby speakers align themselves with their interlocutors for various reasons, for example, to increase their chance of being understood \cite{gil79,cro00}.  If one looks infinitely far into the future, the outcome of a genetic drift process is quantitatively consistent with the fate of New Zealand English \cite{bax09}.  However, this explanation relies on fluctuations to reach the state where all agents have adopted the innovation. Recall from the previous section that the magnitude of these fluctuations decreases with the speech community size. The work of \cite{bax08,bax09} concludes that the timescale of change in such a theory increases with the speech community size in a way that is inconsistent with the rapid pace of language change seen in the example of New Zealand English, where the speech community was large.  In order to see a more rapid change, or to see a directed change, a more powerful asymmetry is needed.

\section{Asymmetry in Social Attitudes}
\label{sec:speaker}

One way to introduce further asymmetry is if speakers have different attitudes towards each others' behaviour.  In particular, in an interaction between speakers $i$ and $j$, there is no particular reason why speaker $i$ should give the same weight to speaker $j$'s utterances as the other way round.  The question now is whether this asymmetry can generate the sustained directed growth of an innovation.

For this to be possible, there must be some correlation between this asymmetry and the set of speakers who initially use the innovation.  To understand why, consider the opposite case where there are no correlations between the influence that a speaker has an whether they initially use the innovation. The average influence of speakers who use the innovation is then, by definition, equal to the average influence of speakers who do not. This averaging out of influence then restores the symmetry between the variants, and thus one would not expect directed changes to arise.

In models of innovation diffusion, some relationship between innovativeness and social influence is typically assumed (albeit with varying degrees of explicitness).  For example, Rogers \cite{rog03} refers to a group of `innovators' who have influence over an `early majority' who, in turn, have influence of a 'late majority' and so on. In the sociolinguistics literature, there has been some discussion of social networks, focussing on the role that strong ties between individuals might play as a mechanism to preserve social norms, and how the number and quality of relationships between in governing how linguistic variation propagates (see e.g.~\cite{mil87,cha03}). Meanwhile, Labov \cite{lab01} and Rogers \cite{rog03} further emphasise the important role played by specific individuals who have influence over other members of a social group when it comes to propagating an innovation. These all imply some sort of asymmetry in social influence.

What we have found when incorporating social asymmetry into a model of language change is that an innovation which is initially used within a small group of influential users can grow in frequency over a sustained period.  However, the shape of the adoption curve depends somewhat on the details of how the social network is configured \cite{bly12}.  This one can see by asking how the average usage frequency of the innovation changes in the presence of social asymmetry.

To this end, let us return to the mathematical model of Sec.~\ref{sec:neutral}, and generalise it to the case of asymmetric interactions.  This we shall do by redefining the quantity $G_{ij}$. Previously, this was the probability that agents $i$ and $j$ interact in a time interval of length $\delta t$.  We now take it to be equal to the probability that agents $i$ and $j$ interact in this time interval \emph{and} that agent $i$ modifies its usage frequency in response to agent $j$'s utterances.  Clearly we no longer require $G_{ij}=G_{ji}$. If agent $i$ seeks to emulate agent $j$ more than the other way round, we will have $G_{ij} > G_{ji}$; otherwise the converse will be true.

In terms of the mathematics, Eqs.~(\ref{eq:xitilde}) to (\ref{eq:xdash1}) are unaffected by this redefinition.  However, the relationship (\ref{eq:cancel}) crucially depended on the symmetry $G_{ij}=G_{ji}$.  This time, we find instead from (\ref{eq:xdash1}) that
\begin{equation}
x' - x = \frac{1-\alpha}{N} \sum_i \left[ \sum_{j\ne i} G_{ij} x_j - \sum_{j \ne i} G_{ij} x_i \right] = \frac{1-\alpha}{2N} \sum_i \sum_{j \ne i} ( G_{ij} - G_{ji} ) (x_j - x_i) 
\end{equation}
where we have twice used the fact that $\sum_i \sum_{j \ne i} f_{ij} = \sum_i \sum_{j \ne i} f_{ji}$ by exchanging indices and reversing the order of summation.

This expression shows that the interaction asymmetry $G_{ij} - G_{ji}$ is crucial in determining the rate of change of the usage frequency $x$.  First of all, we can confirm our intuition that where the language behaviour (encoded here by the differences $x_j-x_i$) is uncorrelated with the interaction asymmetries, the above sum will be of order $1/\sqrt{N}$, and consequently we expect the dynamics to be similar to the case of no asymmetry (see Section~\ref{sec:neutral}).

Second, when $G_{ij} - G_{ji}$ is positive, we see that the usage frequency tends to increase if speaker $j$ is more innovative than speaker $i$ (i.e., if $x_j > x_i$), and it tends to decrease otherwise.  This is to be expected, since we have $G_{ij} > G_{ji}$ when speaker $i$ pays more attention to speaker $j$ than vice versa.  More significantly, this observation has implications for the shape of an adoption curve when the frequency of an innovation is small.

To see this, suppose initially that some speakers are innovators, and have $x_i=1$, whilst the remainder of the speech community are all categorical uses of the existing convention, and have $x_i=0$.  Suppose also that the innovators exert influence over non-innovators.  Then, for this initial condition we have
\begin{equation}
x' - x = \frac{1-\alpha}{N} \sum_{[ij]} (G_{ij} - G_{ji})
\end{equation}
where here the notation $[ij]$ refers to ordered pairs $i,j$ such that speaker $j$ is an innovator and speaker $i$ is not. The statement that the innovators exert influence over non-innovators implies that $G_{ij}>G_{ji}$ for all such pairs.  Hence, $x'-x$ is a strictly positive quantity even with small numbers of innovators: that is, there is some positive rate of growth at low innovation frequencies.  Simulation results \cite{bly12} show that this initial growth can be sufficiently rapid to be inconsistent with an S-shaped adoption curve.  In particular, it was shown that the initial period of slow growth could only be realised under conditions where the size of each successive group in the chain of adopters increased exponentially along the chain \cite{bly12}. As far as we are aware, this does not match any known population structure.

In summary, when one relies on asymmetry in social attitudes to drive the adoption of an innovation, universality in the third (physics) sense does not apply.  The details of the network of social influences matter, at least in terms of the initial shape of the adoption curve. Therefore, this type of asymmetry does not provide a robust explanation for the universal S-shaped trajectory of language change.

\section{Asymmetry in the Variants}

It turns out that a robust explanation for the S-curve is provided by an asymmetry in speakers' attitudes towards the linguistic behaviour itself rather than its users.  To model this, we now introduce an explicit bias $f(x_j)$ into agent $i$'s estimate of its usage frequency among its interlocutors. Specifically, we now take
\begin{equation}
\tilde{x}_i = \frac{\sum_{j\ne i} G_{ij} \left[ x_j + f(x_j) \right]}{\sum_{j \ne i} G_{ij}}
\end{equation}
instead of the unbiased expression (\ref{eq:xitilde}). Whenever the bias $f(x_j)$ is positive, the frequency of the innovation is over-estimated relative to its actual value; likewise when it is negative, the frequency is under-estimated.  We impose two constraints on the form of $f(x_j)$.  First, we insist that it vanishes when $x_j=0$ or when $x_j=1$. This is to be consistent with our approach, in which the innovation process is implicit in the initial condition: if we did not have $f(0)=f(1)=0$, the innovation would be spontaneously recreated if it goes extinct.  We also insist that $0 \le x_j + f(x_j) \le 1$ for all $x_j$, so that it can be interpreted as a frequency in the same way as $x_j$.

If we take this variant-based asymmetry to be the sole asymmetry in the system, we will have $G_{ij} = G_{ji}$, as in Section~\ref{sec:neutral}. Using the above expression for $\tilde{x}_i$ in Eq.~(\ref{eq:xdash1}) we find that
\begin{equation}
x' - x = \frac{1-\alpha}{N} \sum_{i} G_i f(x_i) \;.
\end{equation}
We can now perform the same experiment as in the previous section, where we assign $x_i=1$ to a group of innovators, and $x_i=0$ to a group of conformists and ask for the initial shape of the change trajectory.  Since $f(x_i)=0$ in both cases, we find that $x'=x$, showing that with this initial condition, the frequency of the innovation can change only through a fluctuation.  This is, however, not the same as the fluctuation-driven dynamics that arises when the dynamics are fully symmetric (as described in Section~\ref{sec:neutral}).  There, $x'=x$ no matter what the individual usage frequencies $x_i$ are.  Here, $x'=x$ only if all usage is categorical: as soon as some individuals show variable behaviour (for example, as they start to adopt the innovation), we will have $x'>x$ in the case where the bias acts in favour of the innovation (i.e., if $f(x)>0$).

It is worth emphasising the crucial difference between the initial growth of an innovation arising from speaker-based asymmetry (Section~\ref{sec:speaker}) and the variant-based asymmetry just described. For speaker-based asymmetry, we found that with a small number of categorical innovators, the mean growth rate of the innovation is nonzero, which corresponds to a rapid initial growth. Here, for variant-based asymmetry, we found under the same conditions that the mean growth rate of the innovation vanishes, and so a slower initial growth arises. These expectations were confirmed with an explicit model in \cite{bly12}, which led us to hypothesise that this variant-based asymmetry is a crucial component of language change in real speech communities.

\section{Asymmetry in Variant Frequencies}

The foregoing does not cover all possible asymmetries that might exist in linguistic behaviour.  In particular, one way in which variants could be discriminated is through their frequencies alone, without reference to any aspect of the behaviour itself or association with its users.  This is actually a specific type of variant asymmetry, and as such can be modelled through an appropriate choice of the function $f(x)$ that was introduced in the previous section.

Suppose there are just two variants in competition with each other. Although we will allow the bias $f(x)$ to vary with frequency, we will do so in a way that is symmetric with respect to the variants: that is, the boost applied to a variant with some specific frequency $x_0$ is the same, regardless of which variant this is. In the two-variant case, the two frequencies are $x$ and $1-x$. The symmetry between them implies that the function $f(x)$ must satisfy the constraint
\begin{equation}
f(1-x) = - f(x) \;.
\end{equation}
Again, we will require that $f(0)=f(1)=0$, so that any innovation is innovated only once, and all subsequent adoption of the imitation arises from social interactions.

The simplest functions that satisfy these requirements are $f(x)=0$ and $f(x)=x(1-x)(2x-1)$, which corresponds to boosting a variant if it is a majority variant, and suppressing it if it is in the minority.  This type of frequency boosting, or regularisation, has been observed in a variety of frequency learning experiments, both in the linguistic and non-linguistic domain \cite{hud05,rea09}. A difficulty with this type of model is that there is a threshold problem: low frequency variants face an uphill struggle to reach a frequency of $50\%$, which is needed for the regularisation bias to act in their favour. The presence of noise complicates matters. If the magnitude of any fluctuations is small, this reasoning (based primarily on deterministic considerations) continues to hold. However, when fluctuations are large there is in fact a transition into a regime where the regularisation bias is suppressed, and the usage frequencies fluctuate in the same way as in the fully symmetric case described in Section~\ref{sec:neutral} \cite{rus11}.

This suggests that the main mechanism for propagating an innovation along an S-shaped adoption curve is if the function $f(x)$ is positive for all $x$, thereby providing a systematic bias in favour of the innovation at all frequencies.  However, this raises the question of where this bias comes from.  In \cite{bly12}, it is suggested that speaker-based asymmetry provides a means for speakers to create an association between a group of speakers and a particular linguistic behaviour. Once this happens, a variant-based asymmetry whose origins lie in speaker-based asymmetry may arise. Whether this scenario can be realised spontaneously through local interactions in an agent-based model the subject of a current investigation \cite{jonip}.

Another possibility, raised in \cite{sta14}, is to distinguish between variants not in terms of their current usage frequencies, but according to whether they are increasing or decreasing.  A `momentum-based' bias \cite{gur09} towards further increasing the frequency of a variant which has been increasing in the past could in principle propagate an innovation without appealing to a bias that is based on speaker identity. Again, the question of whether this can arise purely through local interactions between speakers is being investigated with reference to an agent-based model \cite{sta15}.

\section{Discussion}

In this short article, we have explored the various notions of universality in language change. Drawing inspiration from the relationship between symmetry and universality in physics, we have appealed to symmetry as a means to categorise theories for language change. Specifically, we identified the following sources of asymmetry in models of language change: variation in interaction frequencies alone (which corresponds to the theories of accommodation and determinism \cite{tru04}); asymmetry in the degree of influence that speakers have over each other (which correspond to theories based on social network effects, propounded for example by Bloomfield \cite{blo33}, Labov \cite{lab01}, Milroy \cite{mil87} and others); variation in the attitude towards different linguistic variants (which correspond to theories based on prestige and related social factors, advanced for example by Sturtyvant \cite{stu47}, Labov \cite{lab01} and enjoys some prominence among sociolinguists); and finally asymmetry that is based on the usage frequencies of variants (such as regularisation effects \cite{hud05} and momentum-based explanations for change \cite{gur09}).

The key message is that only some of these distinct sources of asymmetry are compatible with the widely-observed (`universal') S-shaped curve for the adoption of an innovation.  We found that a robust model that generates the S-shaped curve can be achieved with a prestige-based explanation (i.e., different attitudes to particular ways of speaking) or potentially with a momentum-based explanation \cite{sta14,sta15}.  In this work, this was determined primarily by investigating the initial rate of growth of an innovation within a fairly general mathematical framework, complementing existing studies that were based on specific simulation models.

A crucial open question that remains is the following. Appealing to symmetries is useful as it allows broad classes of explanations for language change to be excluded with reference only to qualitative features of empirical data. However, this is insufficient to identify a single theory for language change: more than one is compatible with the qualitative data. The challenge then is to distinguish between these remaining theories.  One particular issue with a social prestige type explanation is how the bias towards one linguistic variant over another becomes embedded in the speech community. In \cite{bly12} it was found that a majority of speakers should be positively disposed towards the innovation: how does this positive disposition itself spread through the speech community?  The momentum-based theory of \cite{sta14,sta15} potentially side-steps this issue, since the variants are distinguished by their usage history. If different members of the speech community agree that an innovation is becoming more prevalent, they will all boost the frequency of the same variant. Whilst this is perhaps a more parsimonious theory, that is not in itself sufficient to conclude that it is the more appropriate one. Instead, some independent empirical evidence in favour of a specific explanation is needed. Even better would be to demonstrate that the favoured theory shows greater \emph{quantitative} agreement with empirical data at both the individual and population level. The complexity of human behaviour and social interactions is such that this will be a challenging task, but one where sustained research effort would certainly be worthwhile.

\begin{acknowledgement}
I thank my collaborators Gareth Baxter, Bill Croft, Simon Kirby, Alan McKane, Kenny Smith and Kevin Stadler with whom much of the work outlined here was done.  I also thank Miriam Meyerhoff, Sali Tagliamonte and Peter Trudgill for the sociolinguistic insights that they have shared with me.
\end{acknowledgement}
\bibliographystyle{spphys}
\bibliography{symuni}

\begin{thebibliography}{10}
\providecommand{\url}[1]{{#1}}
\providecommand{\urlprefix}{URL }
\expandafter\ifx\csname urlstyle\endcsname\relax
  \providecommand{\doi}[1]{DOI \discretionary{}{}{}#1}\else
  \providecommand{\doi}{DOI \discretionary{}{}{}\begingroup
  \urlstyle{rm}\Url}\fi

\bibitem{boy85}
W.~Boyd, P.J. Richerson, \emph{Culture and the evolutionary process}
  (University of Chicago Press, Chicago, 1985)

\bibitem{kel94}
R.~Keller, \emph{On language change: The invisible hand in language}
  (Routledge, London, 1994)

\bibitem{cro00}
W.~Croft, \emph{Explaining language change: An evolutionary approach} (Longman,
  Harlow, 2000)

\bibitem{lab01}
W.~Labov, \emph{Principles of lingustic change {II}: Social factors}
  (Blackwell, Oxford, 2001)

\bibitem{wals13}
M.S. Dryer, M.~Haspelmath (eds.), \emph{WALS Online} (Max Planck Institute for
  Evolutionary Anthropology, Leipzig, 2013).
\newblock \urlprefix\url{http://wals.info/}

\bibitem{tag11}
S.A. Tagliamonte, \emph{Variationist Sociolinguistics: Change, Observation,
  Interpretation} (Wiley, New York, 2011)

\bibitem{rog03}
E.M. Rogers, \emph{Diffusion of innovations}, 5th edn. (Free Press, New York,
  2003)

\bibitem{mau11}
L.~Maurits, Representation, information theory and basic word order.
\newblock Ph.D. thesis, University of Adelaide (2011)

\bibitem{oha83}
J.~Ohala, in \emph{The Production of Speech}, ed. by P.F. MacNeilage (Springer,
  New York, 1983), pp. 189--216

\bibitem{cul12}
J.~Culbertson, P.~Smolensky, G.~Legendre, Cognition \textbf{122}, 306 (2012)

\bibitem{rea09}
F.~Reali, T.L. Griffiths, Cognition \textbf{111}, 317 (2009)

\bibitem{cas09}
C.~Castellano, S.~Fortunato, V.~Loreto, Reviews of Modern Physics \textbf{81},
  591 (2009)

\bibitem{hru09}
D.J. Hruschka, M.H. Christiansen, R.A. Blythe, W.~Croft, P.~Heggarty, S.S.
  Mufwene, J.B. Pierrehumbert, S.~Poplack, Trends in Cognitive Sciences
  \textbf{13}, 464 (2009)

\bibitem{smi14}
A.D.M. Smith, Wiley Interdisciplinary Reviews: Cognitive Science \textbf{5},
  281 (2014)

\bibitem{kit05}
C.~Kittel, \emph{Introduction to solid state physics}, 8th edn. (Wiley,
  Hoboken, NJ, 2005)

\bibitem{gol92}
N.~Goldenfeld, \emph{Lectures on Phase Transitions and the Renormalization
  Group} (Addison-Wesley, Reading, MA, 1992)

\bibitem{bly12}
R.A. Blythe, W.~Croft, Language \textbf{88}, 269 (2012)

\bibitem{cha95}
J.K. Chambers, Journal of English Linguistics \textbf{23}, 155 (1995)

\bibitem{pop07}
S.~Poplack, E.~Malvar, Probus \textbf{19}, 121 (2007)

\bibitem{gri09}
A.B. Grieve-Smith, The spread of change in french negation.
\newblock Ph.D. thesis, University of New Mexico (2009)

\bibitem{gre54}
J.H. Greenberg, C.E. Osgood, S.~Saporta, in \emph{Psycholinguistics: A survey
  of theory and research problems}, ed. by C.E. Osgood, T.A. Sebeok (Waverly,
  Baltimore, 1954), pp. 146--63

\bibitem{kro89}
A.S. Kroch, Language Variation and Change \textbf{1}, 199 (1989)

\bibitem{cha02}
J.K. Chambers, in \emph{Handbook of language variation and change}, ed. by J.K.
  Chambers, P.~Trudgill, N.~Schilling-Estes (Blackwell, Oxford, 2002), pp.
  349--72

\bibitem{den03}
D.~Denison, in \emph{Motives for language change} (Cambridge University Press,
  Cambridge, 2003), pp. 54--70

\bibitem{fis30}
R.A. Fisher, \emph{The genetical theory of natural selection} (Clarendon,
  Oxford, 1930)

\bibitem{wri31}
S.~Wright, Genetics \textbf{16}, 97 (1931)

\bibitem{bax08}
G.J. Baxter, R.A. Blythe, A.J. McKane, Physical Review Letters \textbf{101},
  258701 (2008)

\bibitem{bly10}
R.A. Blythe, Journal of Physics A: Mathematical and Theoretical \textbf{43},
  385003 (2010)

\bibitem{tru04}
P.~Trudgill, \emph{New-dialect formation: The inevitability of colonial
  Englishes} (Edinburgh University Press, Edinburgh, 2004)

\bibitem{gil79}
H.~Giles, P.M. Smith, in \emph{Language and Social Psychology}, ed. by
  H.~Giles, R.N. {St Clair} (Basil Blackwell, Oxford, 1979)

\bibitem{bax09}
G.J. Baxter, R.A. Blythe, W.~Croft, A.J. McKane, Language Variation and Change
  \textbf{21}, 257 (2009)

\bibitem{mil87}
L.~Milroy, \emph{Language and Social Networks} (Blackwell, Oxford, 1987)

\bibitem{cha03}
J.K. Chambers, \emph{Sociolinguistic theory : linguistic variation and its
  social significance /} (Blackwell, Oxford, 2003)

\bibitem{hud05}
C.L. {Hudson Kam}, E.L. Newport, Language Learning and Development \textbf{1},
  151 (2005)

\bibitem{rus11}
D.I. Russell, R.A. Blythe, Physical Review Letters \textbf{106}, 165702 (2011)

\bibitem{jonip}
A.~Jones, G.J. Baxter, R.A. Blythe.
\newblock Mechanism for the spontaneous generation of shared linguistic
  preferences.
\newblock in prepration

\bibitem{sta14}
K.~Stadler, R.A. Blythe, K.~Smith, S.~Kirby, in \emph{The Evolution of
  Language}, vol.~10, ed. by E.A. Cartmill, S.~Roberts, H.~Lyn, H.~Cornish
  (2014), vol.~10

\bibitem{gur09}
T.M. Gureckis, R.L. Goldstone, Topics in Cognitive Science \textbf{1}, 651
  (2009)

\bibitem{sta15}
K.~Stadler, R.A. Blythe, K.~Smith, S.~Kirby.
\newblock Momentum in language change: A model of self-actuating {S-shaped}
  curvess.
\newblock under review (2015)

\bibitem{blo33}
L.~Bloomfield, \emph{Language} (Holt, Rinehart and Winston, New York, 1933)

\bibitem{stu47}
E.H. Sturtevant, \emph{An introduction to linguistic science} (Yale University
  Press, New Haven, CT, 1947)

\end{thebibliography}
\end{document}